\documentstyle[12pt]{article}
\input epsf

\newcommand{\be}[1]{\begin{eqnarray} \label{#1} }
\newcommand{\eq}{\end{eqnarray}}

\newcommand{\NPB}[1]{Nucl.\ Phys.\ {\bf B#1}}
\newcommand{\PLB}[1]{Phys.\ Lett.\ {\bf B#1}}

\newcommand{\PRD}[1]{Phys.\ Rev.\ {\bf D#1}}

\def\ci{{\cal{I}}}
\newcommand{\f}{\phi}

\newcommand{\FF}{{\cal F}}

\newcommand{\del}{\partial}
\newcommand{\Del}{\nabla}

\begin{document}
\begin{center}
\vskip3em
{\large\bf Quantum Corrections to Monopoles }

\vskip3em
{G.\ Chalmers,\footnote{E-mail:chalmers@insti.physics.sunysb.edu}
M.\ Ro\v cek,\footnote{E-mail:rocek@insti.physics.sunysb.edu}}\\
\vskip .5em {\it Institute for Theoretical Physics\\
State University of New York\\
Stony Brook, NY 11794-3840 USA\\}
\vskip .5cm

{\normalsize R.\ von Unge\footnote{E-mail:unge@feynman.princeton.edu}\\ \vskip
.5em{\it
Joseph Henry Laboratories\\
Princeton University\\
Princeton, NJ  08544 USA }\\
\vskip2em
}
\end{center}

We summarize our recent work on low energy quantum corrections to monopoles
in $N=2$
supersymmetric Yang-Mills theory.  The details maybe found in \cite{theart}.

Most of our understanding of solitons in quantum field theory comes from
semiclassical
expansion about classical solutions.  However, it was realized long ago
\cite{gj} that in
principle there is an alternative: One can study ``classical'' solutions of
the quantum
effective action.  Of course, in practice, one rarely has access to the
full effective action.
Nevertheless, one can study the solutions to various approximations and get
approximate
information. An example of this procedure is the
Skyrme model where solitons of the QCD low energy effective action (depending
on a mesonic field) are considered to be baryons. The key difference
between that case and our problem is, as explained below, that we know the
{\em full}
non-perturbative as well as perturbative low-energy effective action; this
should of course
make our predictions more reliable.

Three years ago, Seiberg and Witten \cite{sw} started a revolution when
they found the full
nonperturbative low-energy effective action of the massless states in $N=2$
supersymmetric
$SU(2)$ Yang-Mills theory.  This effective action can be gauge
covariantized with repect to
the spontaneously broken $SU(2)$ gauge symmetry (as described below) to
describe the massive
charged vector multiplets of the theory.  Evidence for the correctness of
this procedure is
provided by an interesting strong-coupling phenomenon: the massive charged
vector multiplets
destabilize for certain values of the Higgs vacuum expectation value, and
precisely at these
values the effective Hamiltonian changes sign \cite{lgr}.

Using this low-energy nonperturbative approximation to the full effective
action, we find
that the resulting effective Hamiltonian admits BPS solutions that can be
interpreted as
quantum corrected monopoles and dyons.

In this report, we do not have space to review all of the results of
Seiberg and Witten; a
brief summary of the salient points will have to suffice. It is well known
that $N=2$
Yang-Mills theory with gauge group $SU(2)$ has a classical potential with
valleys; different
vacua are described by the VEV of the scalar fields in the theory, and as
long as the VEV
is nonzero, the gauge group is broken to $U(1)$.  Seiberg and Witten argued
that
nonperturbatively, the valleys remain, but that the vacuum with unbroken
gauge symmetry is
destabilized by quantum effects.  They described the nonperturbative
low-energy (up to
quadratic order in gradients) effective action of the massless $U(1)$
fields in terms of a
single holomorphic function $\FF$\cite{FF}.  They parametrized the space of
vacua, or moduli
space, in terms of a single complex variable $u=<Tr\phi^2>$, where $\phi$
is a complex scalar
Higgs field in the adjoint representation of $SU(2)$. The VEV $a$ of the
scalar field in the
unbroken $U(1)$ is given in terms of $u$ as a certain integral, as is
$a_D\equiv\frac{\del\FF}{\del a}$.  The effective coupling and
$\theta$-angle are
given by
$\tau=\frac\theta{2\pi}+i\frac{4\pi}{g^2}\equiv\frac{\del^2\FF}{(\del
a)^2}$.

As already observed by Seiberg and Witten, the low energy interactions of
the massive charged
supermultiplets in the broken part of the $SU(2)$ gauge group can be
described by simply
covariantizing with respect to the gauge group: $\FF(a)\to
\FF(\sqrt{\phi\cdot\phi})$.  If one
regards this as a Wilsonian effective action, it ceases to become
meaningful when the massive
charged vector multiplets are no longer the lightest massive states in the
theory; however, we
see no compelling reason to take a Wilsonian view, and regard
$\FF(\sqrt{\phi\cdot\phi})$ as
the leading low-energy expansion of the 1-PI generating functional (to this
level in the
derivative expansion, the two appear to coincide \cite{wpi}).  Evidence for
this
interpretation is provided by the observation that the effective action
itself encodes the
information about its range of validity: there is a disk bounded by a {\em
curve of marginal
stability} in the $u$-plane inside of which the charged supermultiplets
become unstable and
vanish from the spectrum.  Precisely on this disk, the effective action for
the charged vector
multiplets breaks down by changing sign \cite{lgr}.

The bosonic part of the low-energy effective Hamiltonian that we compute
from this effective
action has a bulk as well as a boundary contribution $H=H_0+H_{\rm top}$:

\be{almostBPS}
H_0  =
{1\over 8\pi} & {\ci}m \int d^3x \FF_{AB} \Bigl\{
 (B^A_i +iE^A_i+\sqrt{2} e^{i\alpha} \nabla_i\f^A)
 (B^B_i-iE^B_i+\sqrt{2} e^{-i\alpha} \nabla_i {\bar\f}^B) \Bigl\}  \ .
\eq
Here $\FF_{AB}\equiv\frac{\del^2\FF}{\del\phi^A\del\phi^B}$, and $B$ and
$E$ are the
nonabelian magnetic and electric fields.
A constant phase $e^{i\alpha}$ has been introduced and will be given
explicitly below. It might appear that $e^{i\alpha}$ could be absorbed
into the complex field $\f$.  However, the asymptotic phase of $\f$
has an independent significance: it determines the $\theta$-angle of the
vacuum. Note that $H_0$ is positive whenever ${\ci}m \FF_{AB}\geq 0$; hence
the BPS equations
for the general monopole and dyonic states is

\be{BPS}
B_j^{B}+iE_j^{B}+e^{i\alpha} \sqrt{2} \Del_j\f^{B} =0  \ .
\eq
These equations have the same form as in the classical
Yang-Mills-Higgs action. Nevertheless, as we explain below, there are
significant
quantum corrections.

The energy associated with the boundary terms is

\be{mass}
H_{top} = -\sqrt{2} \int d{\vec\Sigma} \cdot ~\Bigl\{ {\vec\Pi}_A
{\ci}m ( e^{i\alpha} \f^A) + {1\over 4\pi} {\vec B}^A {\ci}m
( e^{i\alpha} \FF_A ) \Bigr\} \ .
\eq
Here $\Pi$ is the momentum conjugate to the nonabelian vector potential,
and generates gauge
transformations. After some calculation, we find that for
\be{alpha}
e^{i\alpha}= i \frac{\bar Z}{\vert Z\vert} \quad\quad Z=n_e a+n_m a_D \ ,
\eq
where $Z$ is the central charge of the $N=2$ superalgebra,  we
obtain the BPS bound for the total energy
\be{bound}
E \geq \sqrt{2} \vert Z\vert \ .
\eq
When the quantum corrected monopole/dyon solutions
satisfy the {\em  usual} BPS equations, the bound is saturated.

We may next consider a radial ansatz, and study the structure of the
solutions. Conceptually,
we imagine that we are finding a solution $u(r)$, where $u$ is the modular
parameter described
above, and from it determine the $\phi(r)$, $B(r)$, and $E(r)$. A key
change from the classical
solution arises when we consider the appropriate boundary conditions for
the problem. In the
classical theory, one imposes regularity everywhere in space, and
consequently, the
expectation value of the Higgs field vanishes at the origin.  Seiberg's and
Witten's
representation of the VEV $a(u)$ can {\em never} vanish (we call this
excluded region the
hard core of the monopole).  However, this does not lead to a
contradiction: the low-energy
approximation is expected to break down when derivatives become large,
which happens at the
core of the monopole; consequently, we cannot impose a boundary condition
at the origin.  This
introduces an extra parameter into our solution which merely expresses our
ignorance of what
is happening inside the core of the soliton. Examination of how this
parameter enters reveals
that it has a simple physical significance: it determines the effective
contribution of the
massive charged vectors to the soliton.

A more fundmental change in the solution arises because the gauge theory
that we are
considering is asymtoptically free, and undergoes dimensional
transmutation: the coupling and
$\theta$-angle are determined by the value of the Higgs field.
Consequently, $\tau$ becomes
effectively spatially dependent $\tau\to\tau(r)$, as does the central
charge $Z\to Z(r)$. This
has a dramatic effect: charge of the dyon depends of the {\em local} value
of $\tau(r)$, and
hence, the electric field has direct quantum corrections: it is no longer a
simple duality
rotation of the magnetic field.  This is consistent with the breaking of
the continuous
duality group to a discrete subgroup.

The solution that we find is characterized by the condition that the phase
of the local
central charge $Z(r)$ is constant throughout space.  We may follow lines of
constant $arg(Z)$
in the $u$-plane or the $\tau$-plane; these are shown in Figure 1. A
solution should be
thought of starting at some point and following a line to the curve of
marginal stability;
beyond this curve, the solution cannot be trusted, as the massive vector
multiplets are no
longer in the spectrum; this defines a larger ``soft'' core fo the
monopole. We note that for
approximately half the $u$-plane, the solutions all run to one fixed point
$u=1$. Seiberg
and Witten argued that at this point the monopoles are becoming massless;
notice that we
cannot describe such monopoles as quantum corrected classical solutions,
since massless
monopoles have a core that fills all space.

\begin{figure}[hbt]
  \begin{center}

\mbox{\epsfxsize=8.0cm \epsfysize=9.6cm \epsfbox[0 0 471
500]{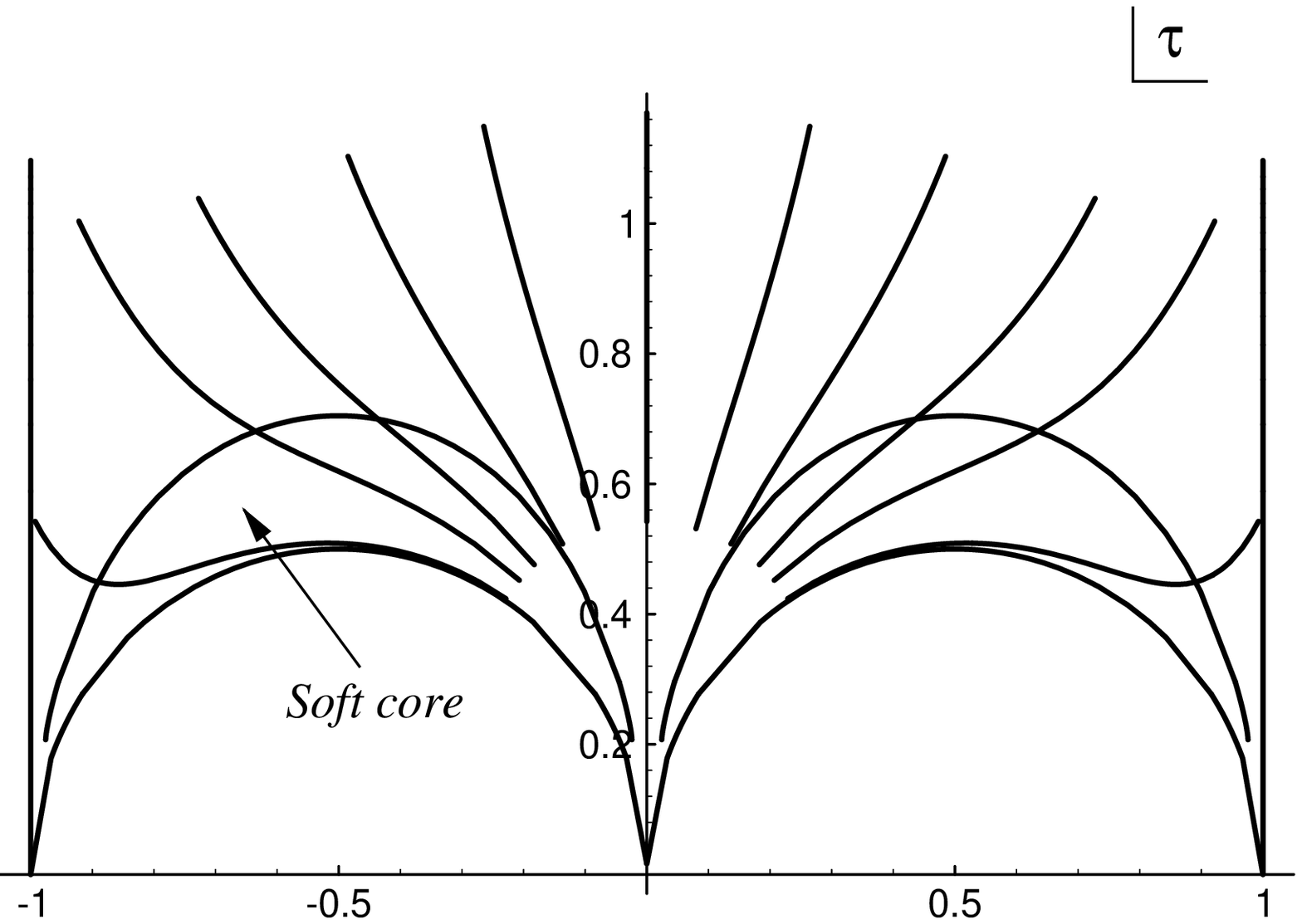}}\hskip .8cm
\mbox{\epsfxsize=7.2cm \epsfysize=9.6cm \epsfbox{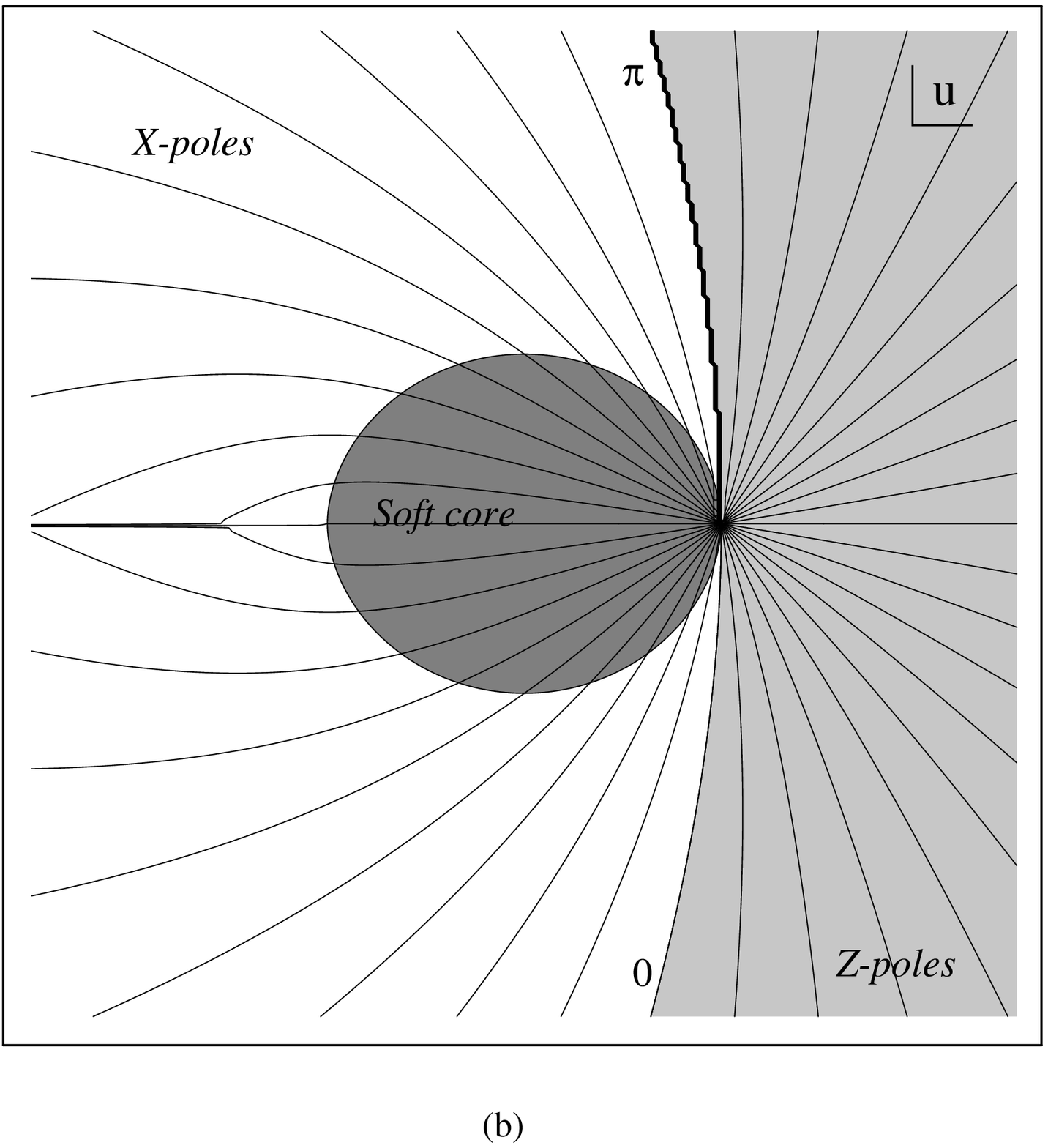}}
  \end{center}
\vskip -.5cm
  \caption{ (a) Some lines of constant $Z(r)$-phase in
the $\tau$-plane. (b) Lines of constant $Z(r)$-phase in the $u$-plane are
shown for
  $\pi/20$ increments.  }\label{manylines}
 \end{figure}

\eject

\end{document}